\documentclass[preprint,prc,showpacs,preprintnumbers,amsmath,amssymb,floatfix]{revtex4}
\usepackage{amsmath}
\usepackage{graphicx}
\usepackage{dcolumn}
\usepackage{bm}
\usepackage{ulem} 
\usepackage{epstopdf}
\usepackage{epsfig}
\usepackage{float}
\usepackage{subfigure}
\usepackage{slashed}

\newcommand{\nc}{\newcommand}       
\nc{\vc}[1] {\mbox{\boldmath $#1$}} 
\nc{\del}       {\partial}              
\nc{\bra}       {\langle}               
\nc{\ket}       {\rangle}               
\nc{\bras}[1]   {\langle #1|}           
\nc{\kets}[1]   {|#1\rangle}            
\nc{\mapleft}[1]{           
 \smash{\mathop{\,          %
  \hbox to 1.5cm{\rightarrowfill}\, }\limits_{#1}}}
\nc{\beq}     {\begin{eqnarray}} \nc{\eeq}    {\end{eqnarray}}
\nc{\nn}      {\\\nonumber} \nc{\vs}      {\vspace{-0.275cm}}
\nc{\fra}    {\frac{1}{2}}
\nc{\mb}        {\mathbf}

\begin{document}


\title{The properties of nuclear matter with lattice $NN$ potential in relativistic Brueckner-Hartree-Fock theory }

\author{Jinniu Hu}
\email{hujinniu@nankai.edu.cn}
\affiliation{School of Physics, Nankai University, Tianjin 300071, China}
\author{Hiroshi Toki}
\email{toki@rcnp.osaka-u.ac.jp}
\affiliation{Research Center for Nuclear Physics (RCNP), Osaka University, Ibaraki, Osaka 567-0047, Japan}
\author{Hong Shen}
\email{songtc@nankai.edu.cn}
\affiliation{School of Physics, Nankai University, Tianjin 300071, China}

\date{\today}

\begin{abstract}
We study the properties of nuclear matter with lattice nucleon-nucleon ($NN$) potential in the relativistic Brueckner-Hartree-Fock (RBHF) theory. To use this potential in such a microscopic many-body theory, we firstly have to construct a one-boson-exchange potential (OBEP) based on the latest lattice $NN$ potential. Three mesons, pion, $\sigma$ meson, and $\omega$ meson, are considered. Their coupling constants and cut-off momenta are determined by fitting the on-shell behaviors and phase shifts of the lattice force, respectively. Therefore, we obtain two parameter sets of the OBEP potential (named as LOBEP1 and LOBEP2) with these two fitting ways. We calculate the properties of symmetric and pure neutron matter with LOBEP1 and LOBEP2. In non-relativistic Brueckner-Hartree-Fock case, the binding energies of symmetric nuclear matter are around $-3$ and $-5$ MeV at saturation densities, while it becomes $-8$ and $-12$ MeV in relativistic framework with $^1S_0,~^3S_1,$ and $^3D_1$ channels using our two parameter sets. For the pure neutron matter, the equations of state in non-relativistic and relativistic cases are very similar due to only consideration $^1S_0$ channel with isospin $T=1$ case.
\end{abstract}

\keywords{Lattice $NN$ force \sep Relativistic Brueckner-Hartree-Fock theory \sep Nuclear matter}

\maketitle

\section{Introduction}

It is an extremely challenging problem to solve the nuclear many-body problem directly from the quantum chromodynamics (QCD) theory in modern nuclear physics. Since the beginning of this century, we anticipate the dawn of treating such a problem from lattice QCD calculation. Hatsuda {\it{et al.}} (HAL collaboration) extracted the nucleon-nucleon ($NN$) potential based on the imaginary-time  Nambu-Bethe-Salpeter (NBS) wave functions in lattice QCD \cite{ishii07, aoki10, aoki12}. The nuclear force is taken from the zero-strangeness sector of the octet-baryon potentials in the flavor-$SU(3)$ limit calculated on the lattice, where the renormalization group improved Iwasaki gauge action and the nonperturbatively improved Wilson quark action were employed on a $32^3\times32$ lattice with the lattice spacing $a=0.121(2)$ fm.  The pion masses are ranging at five values between $468.6$\,MeV to $1161.0$\,MeV\cite{inoue11, inoue12}. These five potentials were able to describe the basic characters of $NN$ potential, such as a strong repulsive force at short distance for the central interaction and an attractive tensor force at intermediate distance.

Recently, Inoue {\it{et al.}} applied such lattice $NN$ potentials on the study of nuclear many-body system from nuclear matter to finite nuclei \cite{inoue13, inoue15}. They obtained a saturation point ($\rho_0=0.414$ fm$^{-3}$, $E/A=-5.4$\,MeV) for symmetric nuclear matter with the lightest quark mass ($m_\pi=468.6$\,MeV, $M_N=1161$\,MeV) using a powerful microscopic nuclear many-body theory, Brueckner-Hartree-Fock (BHF) theory \cite{baldo12}, which can deal with the short-range central and intermediate tensor forces properly. In this case, the maximum mass of the neutron star is found to be 0.53 times the solar mass. The calculated result is far from the empirical saturation property of symmetric nuclear matter, $\rho_0=0.16$ fm$^{-3}$, $E/A=-16$\,MeV, but it still shows a possibility of {\it{ab initio}} calculation of nuclear many-body problem from QCD theory. Furthermore, the light doubly magic nuclei, $^{16}$O and $^{40}$Ca, were also investigated with the BHF theory in harmonic-oscillator basis~\cite{inoue15}. The binding energies per particle for $^{16}$O and $^{40}$Ca are $-2.17$\,MeV and $-2.82$\,MeV, respectively. These works help us greatly understand the connection between lattice QCD and the ground states of nuclear many-body system.

The relativistic Brueckner-Hartree-Fock (RBHF) theory, as a relativistic version of the BHF theory, is able to describe the saturation property of symmetric nuclear matter successfully, through taking the medium effect into the $NN$ potential \cite{brockmann90}. It also can explain the spin-orbit force naturally by the scalar and vector potentials using Dirac equation. The relativistic effect can provide a repulsive contribution to the binding energy through a Z-graph process of nucleon-antinucleon excitation \cite{machleidt89}, instead of a phenomenological three-body force used in the BHF theory. Therefore, in this work, we would like to study the properties of nuclear matter using the lattice $NN$ potential with the RBHF theory. We provide the necessary formula of the RBHF theory in Sect.~II and discussions of our results in Sect.~III, before drawing conclusions in Sect.~IV.
\section{Relativistic Brueckner-Hartree-Fock theory}
In this section, we will show the basic framework of RBHF theory. The effective interaction, $G$-matrix in RBHF theory can be written as the Bethe-Goldstone equation \cite{brockmann90}
\beq\label{ten}
G_{ij}(\mathbf P;\mathbf k, \mathbf k' )=V^*_{ij}(\mathbf k, \mathbf k')+\int\frac{d\mathbf{q}}{(2\pi)^3}V^*_{ij}(\mathbf k, \mathbf q)\frac{Q_{ij}(\mathbf q, \mathbf P)G_{ij}(\mathbf P; \mathbf q, \mathbf k')}{2E^*(\mathbf P/2+\mathbf k')-2E^*(\mathbf P/2+\mathbf q)},
\eeq
where $\mathbf P$ is the c.m. momentum, $\mathbf k,~\mathbf q$ and $\mathbf k'$ are the initial, intermediate, and final relative momenta, respectively, for two particles in nuclear medium. $Q$ is the Pauli operator projecting onto unoccupied states. We can solve this integral equation by the matrix inverse method. Furthermore, $E^*$ is the single-particle energy of nucleon, in nuclear matter, which can be defined as
\beq\label{se}
E^*_i(p)=T_i(p)+U_i(p),
\eeq
where $T_i(p)$ is kinetic energy and $U_i(p)$ is a single-particle potential related with $G$-matrix,
\beq\label{su}
U_i(p)=\bra p|U_i|p\ket=\Re[\sum_{q\leq k^n_F}\bra pq|G_{in}|pq-qp\ket+\sum_{q\leq k^p_F}\bra pq|G_{ip}|pq-qp\ket],
\eeq
where $|p\ket$ and $|q\ket$ are nucleon states including single-particle momenta, spin, and isospin index. The propagator in Eq.(\ref{ten}) depends on the single-particle potential, $U_i~(i=n,p)$ through the single-particle energy, Eq.(\ref{se}). Consequently, the determination of $G$-matrix depends on the choice of $U_i$. In many-body problem, a nucleon in the nuclear medium can be viewed as a 'bare' nucleon that is 'dressed' as a consequence of its effective two-body interactions with the other nucleons. Such a 'dressed' nucleon state should satisfy the Dirac equation in nuclear matter,
\beq
(\slashed{p}_i-m_i-\Sigma_i(p))u_i(\mb p,s)=0,
\eeq
where, $\Sigma_i(p)$ is the relativistic self-energy of nucleon and $i=n/p$ for neutron/proton. As symmetry required, the self-energy must have the general Lorentz structure
\beq
\Sigma_i(p)=U_{S,i}(p)+\gamma_0U^0_{V,i}(p)-\bm\gamma\cdot\mb pU_{V,i}(p),
\eeq
where $U_{S,i}$ and $U_{V,i}$ are an attractive scalar field and a repulsive vector field, respectively, and $U^0_{V,i}$ is the time component of the vector field. It has been shown that $U_{V,i}$ is much smaller than $U_{S,i}$ and $U^0_{V,i}$ in Ref.~\cite{brockmann90}. Thus we can write
\beq\label{sea}
\Sigma_i(p)\approx U_{S,i}(p)+\gamma_0U^0_{V,i}(p).
\eeq
Therefore, we can obtain the positive energy solution in above Dirac equation,
\beq\label{ds}
u_i(\mathbf p,s)=\left(\frac{m^*_i+E^*_i(p)}{2m^*_i}\right)^{1/2}\left[\large{\begin{array}{c}1\\ \frac{\bm\sigma\cdot\mathbf p}{m^*_i+E^*_{i}(p)}\end{array}}\right]\chi(s),
\eeq
where $\chi(s)$ is a Pauli spinor, $m^*_i$ the effective nucleon mass and $E^*_i(p)$ effective single-particle energy in relativistic framework,
\beq
m^*_i(p)=m_i+U_{S,i}(p)
\eeq
and
\beq
E^*_i(p)=(m^{*2}_i+{\mb p}^2_i)^{\frac{1}{2}}.
\eeq

It is formally identical to a free-space spinor, but with $m_i$ replaced by $m^*_i$. Now, from Eq.(\ref{sea}), the single-particle potential becomes
\beq
U_i(p)=\frac{m^*_i}{E^*_i}\bra p|U_{S,i}+\gamma^0U^0_{V,i}|p\ket,
\eeq
where $|p\ket$ is a Dirac spinor as shown in Eq.(\ref{ds}) and $\bra p|$ is the conjugate spinor. $U_{S,i}$ and $U^0_{V,i}$ are momentum dependent, which can be taken as the constants in a fixed density as an approximation~\cite{brockmann90}. Thus, the single-particle potential becomes,
\beq
U_i(p)=\frac{m^*_i}{E^*_i(p)}U_{S,i}+U^0_{V,i}.
\eeq
One can easily see that this equation could be parameterized in terms of two constants, $U_{S,i}$ and $U^0_{V,i}$. Therefore, in the actually numerical procedure, starting from some initial values of $m^*_i$ and $U^0_{V,i}$, the $G$-matrix equation is solved and a first approximation for $U_i(p)$ is then obtained. This solution is again parameterized in terms of a new set of constants, and the calculation are repeated until the convergence is reached. The energy per neutron or proton in nuclear matter can be calculated through evaluating the expectation value of $G$-matrix with relativistic Hartree-Fock wave functions.
\section{Results and Discussion}

The $NN$ potential used in the RBHF theory should be the one described by quantum field theory with spinor structure in order to take into account the nuclear medium effect. Therefore, at the beginning, a one-boson-exchange potential (OBEP) should be constructed based on the present lattice $NN$ potential which is represented in a relative coordinate space. In this work, we only consider the lattice $NN$ potential with lightest quark mass ($m_\pi=468.6$\,MeV, $M_N=1161$\,MeV) worked out by Inoue {\it{et al.}}, which is closest to the pion physical value and generates the most attractive binding energy for nuclear matter and finite nuclear system in the BHF theory \cite{inoue13}.

\begin{figure}[!htb]
\centering
\includegraphics[bb=0 250 350 900, width=0.50\textwidth]{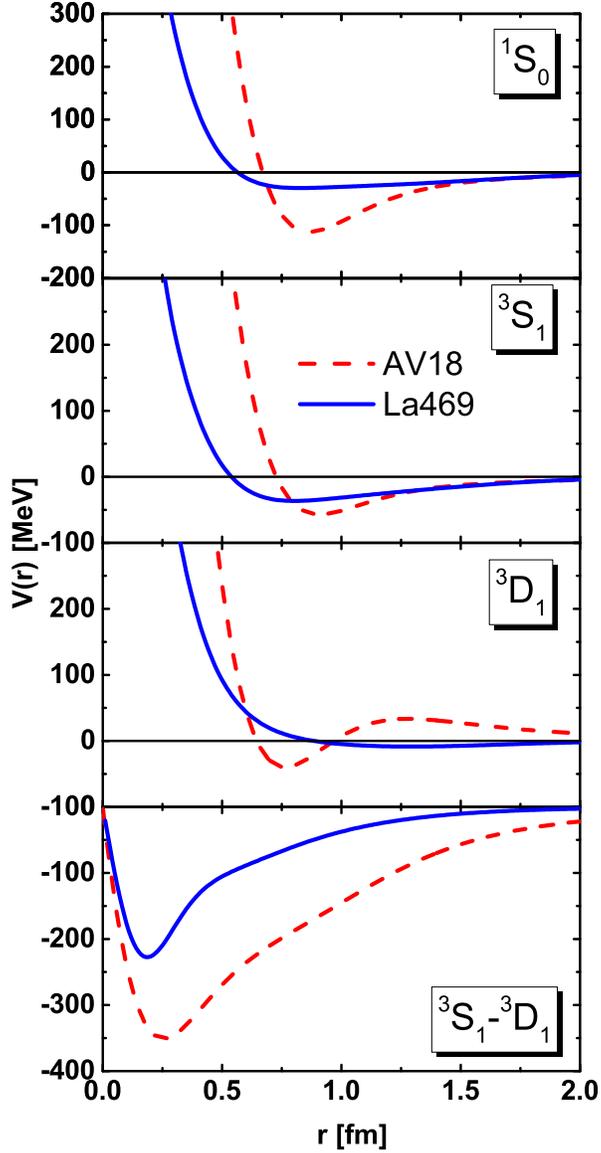}
\caption{The lattice La469 potential in various channels is compared with AV18 potential in the coordinate space. The solid curves represent the La469 potential, where the pion mass is $m_\pi=468.6$\,MeV, while the dashed curves the AV18 potential.}
\label{lar}
\end{figure}

In Fig.\,\ref{lar}, we plot the $NN$ potential from the lattice QCD calculation for various channels in coordinate space, using the potential of the pion mass $m_\pi=468.6$\,MeV (La469). There are only four channels in lattice force, $^1S_0,~^3S_1,~^3S_1-{^3D_1},$ and $^3D_1$ until now with partial waves $L=0, ~2$. The $S$ channel corresponds to the central force, while the $^3S_1-{^3D_1}$ channel to the tensor force. We also compare these potentials with a high precision charge-dependent realistic $NN$ potential obtained by fitting the phase shifts of $NN$ scattering data, AV18 potential \cite{wiringa95}. We can find that the behaviors of lattice potential are similar with the AV18 potential.  At short distance in the $^1S_0$ and $^3S_1$  channels, there are strong repulsive cores, while in the $^3S_1-^3D_1$ channel, an attractive region appears in intermediate distance.  The largest difference between these two potentials is that the attractive magnitude of the lattice QCD is much smaller than the one of AV18 potential. This should arise from the fact that the present pion mass is still far from its physical value, $m_\pi=137$\,MeV.

Usually, the potential written in the momentum space is more convenient for the calculation of nuclear matter. Therefore, we would like to transform the present La469 potential into the momentum space by Fourier transformation:
\beq
&&\bra L'STJM|V(k',k)|LSTJM\ket\nn
&=&\frac{2}{\pi}\int r^2drr'^2dr' j_{L'}(k'r')\bra L'STJM|V(r',r)|LSTJM\ket j_L(kr),
\eeq
where $j_L$ is the $L$-order spherical bessel function.

\begin{figure}[!htb]
\centering
\includegraphics[bb=0 250 350 900, width=0.50\textwidth]{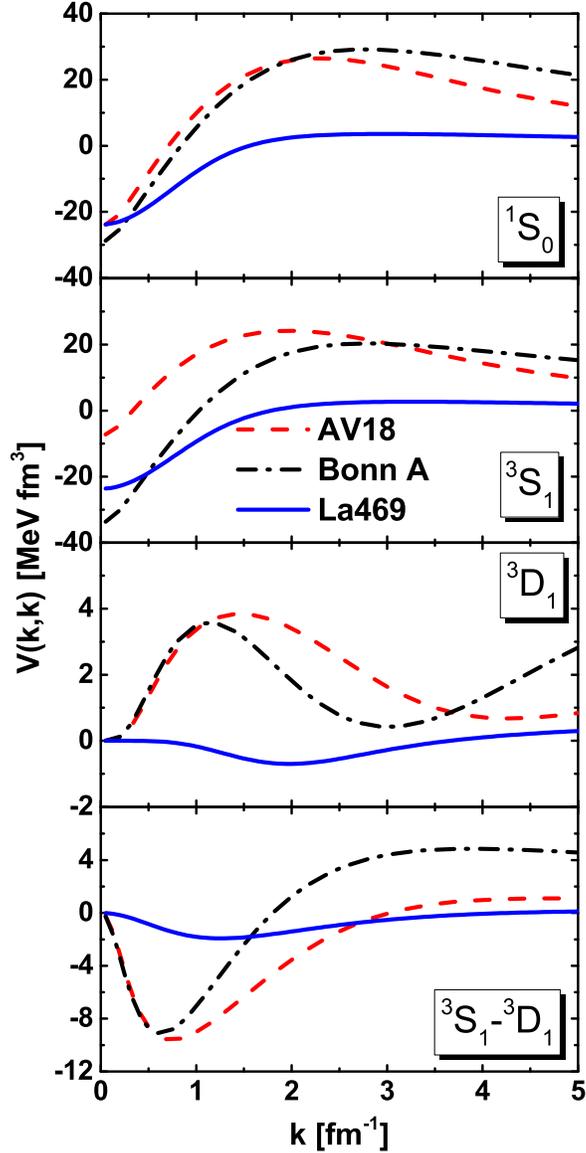}
\caption{The $NN$ on-shell matrix elements from La469 potential, AV18 potential and Bonn A potential for various channels in momentum space. The solid curves represent the La469 potential, the dashed ones the AV18 potential, and dotted-dashed ones the Bonn A potential.}
\label{lam}
\end{figure}

We show the on-shell ($k=k'$) matrix elements of $NN$ potentials of La469 potential together with the AV18 potential for $^1S_0,~^3S_1,~^3S_1-{^3D_1},$ and $^3D_1$ channels in momentum space in Fig.\ref{lam}. Here, we include one more realistic $NN$ potentials, Bonn A potential \cite{machleidt89}, which is the $NN$ potential constructed by one-boson exchange potential (OBEP) with six mesons, $\sigma,~\omega,~\pi,~\delta,~\rho,$ and $\eta$. The behavior of on-shell matrix elements in La469 potential is similar to those of AV18 potential and Bonn potential for $^1S_0,~^3S_1,$ and $^3S_1-{^3D_1}$ channels. However, there is an attractive force at low momentum in the $^3D_1$ channel of La469 potential, while it is repulsive in AV18 potential and Bonn potential.

It is necessary to use a OBEP in the RBHF theory to discuss the relativistic effect in nuclear matter. So, we try to construct a OBEP by using the La469 potential. Three mesons, pion, $\sigma$, and $\omega$ mesons are considered in the present OBEP. Pion can provide the tensor force and long range part of $NN$ potential, while the $\sigma$ and $\omega$ mesons provide middle-range and short-range contributions to $NN$ potential, respectively. In this case, the Lagrangian of mesons coupling with nucleon can be written as,
\beq
\mathcal{L}_\sigma&=&+g_\sigma\overline{\psi}\sigma\psi,\nn
\mathcal{L}_\omega&=&-g_\omega\overline{\psi}\gamma^\mu\omega_\mu\psi,\nn
\mathcal{L}_{\pi}&=&-\frac{f_{\pi}}{m_{\pi}}\overline{\psi}\gamma_5\gamma^\mu\partial_\mu\vec\pi\cdot\vec\tau\psi,
\eeq
where, the pseudovector coupling between pion and nucleon is adopted and $g_\pi/2M_N=f_\pi/m_\pi$.

To discuss the influences of on-shell and off-shell behaviours of lattice potential, two parameter sets of OBEP will be fitted. In the first parameter set, we fit the parameters in OBEP, the coupling constants between mesons and nucleon and the cut-off momenta, through the on-shell matrix elements of La469 potential, and obtain LOBEP1. The second parameter set, LOBEP2, is obtained by fitting the phase shift of La469 potential. In above two fittings, the parameters are determined by the data of La469 potential in $^1S_0,~^3S_1,$ and $^3S_1-{^3D_1}$ channels to avoid anomalous behaviors in $^3D_1$ channel of La469 potential seen in comparison with AV18 and Bonn potentials.

\begin{figure}[!htb]
\centering
\includegraphics[bb=0 135 350 700, width=0.50\textwidth]{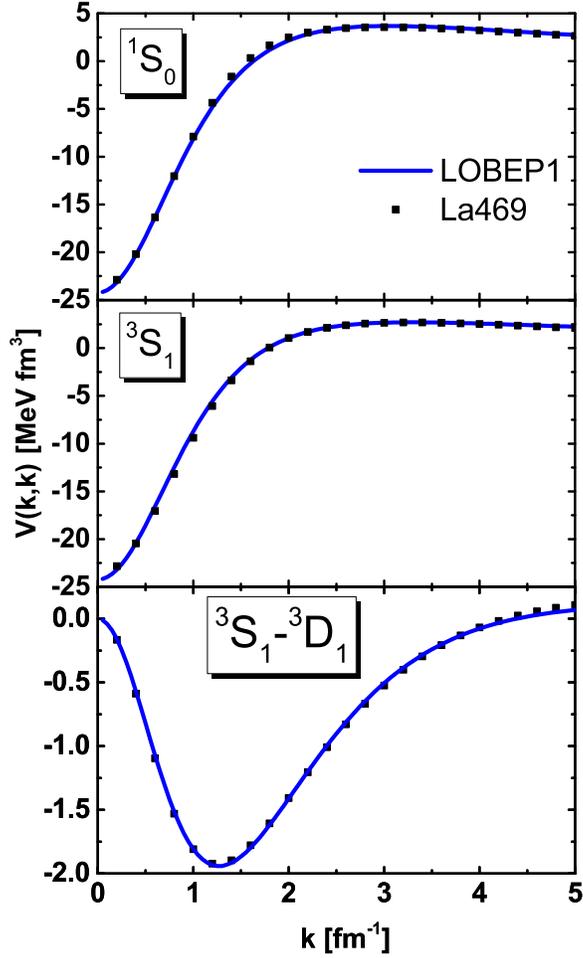}
\caption{The on-shell matrix elements of La469 potential and fitted LOBEP1 potential  for various channels in momentum space. The solid curves represent the LOBEP1 potential, while the symbols are the La469 potential.}
\label{obpe1}
\end{figure}

In Fig.\,\ref{obpe1}, we plot the on-shell matrix elements of OBEP by fitting the La469 potential (LOBEP1) for $^1S_0,~^3S_1$, and $^3S_1-{^3D_1}$ channels in momentum space.  We find that the LOBEP1 can describe the on-shell behaviors of La469 potential very well.

\begin{figure}[!htb]
\centering
\includegraphics[bb=0 135 350 700, width=0.50\textwidth]{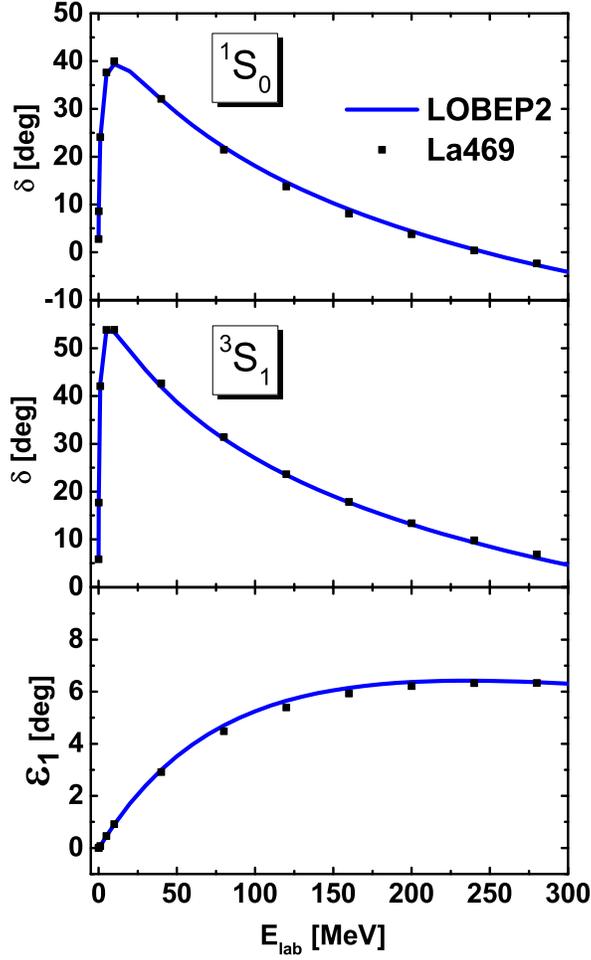}
\caption{The phase shifts of La469 potential and fitted LOBEP2 potential at different channels. The solid curves represent the phase shifts of LOBEP2, while the symbols are the ones of La469 potential.}
\label{obpe2}
\end{figure}

The La469 potential is extracted from the lattice results through Schroedinger equation in the non-relativistic framework. Therefore, the phase shifts of La469 potential is calculated within the non-relativistic propagator, while the OBEP in the RBHF model should be obtained in the relativistic form following the idea in Ref.\cite{brockmann90}. We use the Thompson equation to work out the phase shifts to fix LOBEP2 potential \cite{machleidt93}. Since the phase shifts are the observable quantities, they should be independent of frameworks. The phase shifts of La469 potential and the fitted LOBEP2 potentials, are given in Fig. \ref{obpe2} for $^1S_0,~^3S_1$, and $^3S_1-{^3D_1}$ channels. In the third panel, $\varepsilon_1$ is the mixing parameter of $^3S_1-{^3D_1}$ coupled states. These phase shifts of LOBEP2 potential can describe La469 potential very well, where $\chi^2/N_\text{data}\sim0.2$ up to the laboratory energy of $E_\text{lab}=300$\,MeV.

We tabulate the meson-nucleon coupling constants and cut-off momenta of LOBEP1 and LOBEP2 potentials in Table\,\ref{set} and compare them with the corresponding values in Bonn\,A potential. The masses of pion, $\omega$ meson, and nucleon have been calculated in lattice QCD. The $\sigma$ meson mass should be fitted in this work. These fitting parameters look reasonable comparing with Bonn\,A potential obtained by the $NN$ scattering data.
\begin{table*}[!htb]
\begin{center}
\scriptsize
\begin{tabular}{c c c c |c c c c|c c c c}
\hline\hline    {Force} &    {             }        & {LOBEP1}             & {        }                &{~~}&{        }             & {    LOBEP2    }    &  {}                       &{~~}&{        }             & {    Bonn A    }    &  {}     \\
                {     } &    {$m_\alpha$ (MeV)}     & {$g^2_\alpha/4\pi$}  & { $\Lambda_\alpha$ (MeV) }&{~~}&{$m_\alpha$ (MeV)}     & {$g^2_\alpha/4\pi$} & { $\Lambda_\alpha$ (MeV) }&{~~}&{$m_\alpha$ (MeV)}     & {$g^2_\alpha/4\pi$} & { $\Lambda_\alpha$ (MeV) } \\
\hline          {$\pi$} &    {$468.6$         }     & {$19.70$          }  & { $814.02$  }             &{~~}&{$468.6  $      }      & {$17.00        $}    & { $902.78$ }                &{~~}&{$138.03  $      }     & {$14.9        $}    & { $1050$ } \\
             {$\sigma$} &    {$492.6$        }      & {$7.97$}             & { $718.95$  }             &{~~}&{$491.9   $      }     & {$8.56$      }      & { $699.27$  }               &{~~}&{$550.0   $      }     & {$8.3141$      }    & { $2000$  }\\
             {$\omega$} &    {$829.2$         }     & {$14.03$}            & { $1126.84$              }&{~~}&{$829.1   $      }     & {$12.76         $}  & { $1129.03$ }                &{~~}&{$782.6   $      }     & {$20.0         $}   & { $1500$ }\\
\hline\hline
\end{tabular} \end{center}
\caption{The meson parameters of LOBEP1, LOBEP2 and Bonn\,A potential~\cite{machleidt89}. In LOBEP1 and LOBEP2 potentials, the nucleon masses are taken as $M_N=1161.0$\,MeV from lattice calculation.}
\label{set}
\end{table*}

\begin{figure}[!htb]
\centering
\includegraphics[bb=0 50 330 470, width=0.50\textwidth]{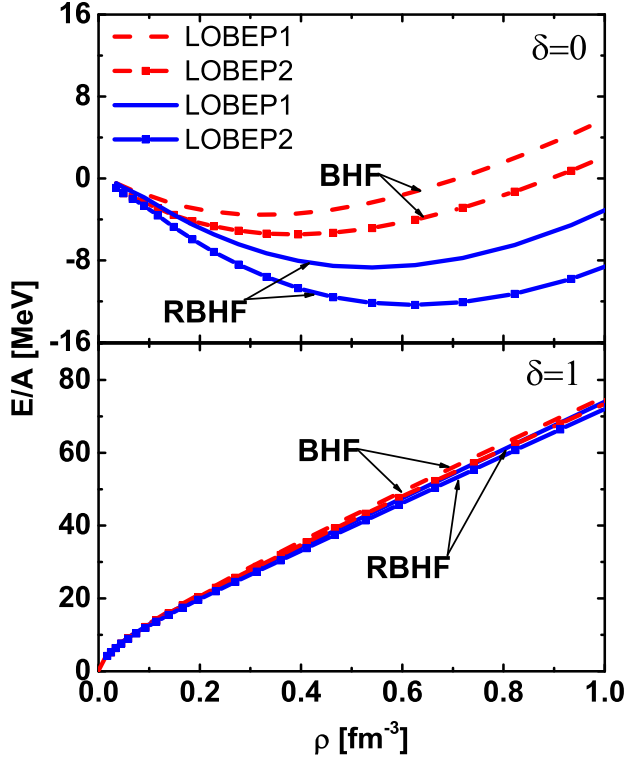}
\caption{The EOSs of symmetric nuclear matter ($\delta=0$) and pure neutron matter ($\delta=1$) with the BHF and RBHF theories for LOBEP1 and LOBEP2 forces. The BHF results are shown with the dashed curve. The RBHF results are given in the solid line.}
\label{EOSSM}
\end{figure}

Now, with the LOBEP1 and LOBEP2 potentials, the equation of state (EOS) of nuclear matter can be calculated with the BHF theory and RBHF theory described in the Methods section as following \cite{brockmann90}, taking only $^1S_0,~^3S_1,~^3S_1-{^3D_1},$ and $^3D_1$ channels. The EOSs of symmetric nuclear matter ($\delta=\frac{N-Z}{A}=0$) in the BHF and RBHF theories are given in the upper panel of Fig.\,\ref{EOSSM}. In the BHF theory, the binding energy is $E/A=-3.56$\,MeV at saturation density $\rho=0.33$ fm$^{-3}$ for the LOBEP1 potential. The saturation properties will be changed as $E/A=-5.47$\,MeV at saturation density $\rho=0.39$ fm$^{-3}$ for LOBEP2 potential, which are consistent with the results by Inoue {\it{et al.}} \cite{inoue13}.  The saturation properties of symmetric nuclear matter are $E/A=-8.67$\,MeV at $\rho=0.54$ fm$^{-3}$ in the RBHF theory with LOBEP1 potential, while they are $E/A=-12.34$\,MeV at $\rho=0.63$ fm$^{-3}$ for LOBEP2 potential. It looks that the non-relativistic case provides more repulsive effect with the present lattice $NN$ potential, while with Bonn potential, the RBHF theory is more repulsive \cite{brockmann90}. But we should remember that there are only $L=0,~2$ channels available in the present lattice potential. We do not have the data of lattice potential at $L=1$ channels.

Furthermore, we also calculate the EOSs of pure neutron matter with LOBEP1 and LOBEP2 potentials in BHF and RBHF theories shown in the lower panel of Fig.\,\ref{EOSSM}. They are almost identical, since in pure neutron matter there is only one contribution from $^1S_0$ channel from lattice $NN$ potential with isospin $T=1$ channel.

\begin{figure}[!htb]
\centering
\includegraphics[bb=0 0 320 240, width=0.50\textwidth]{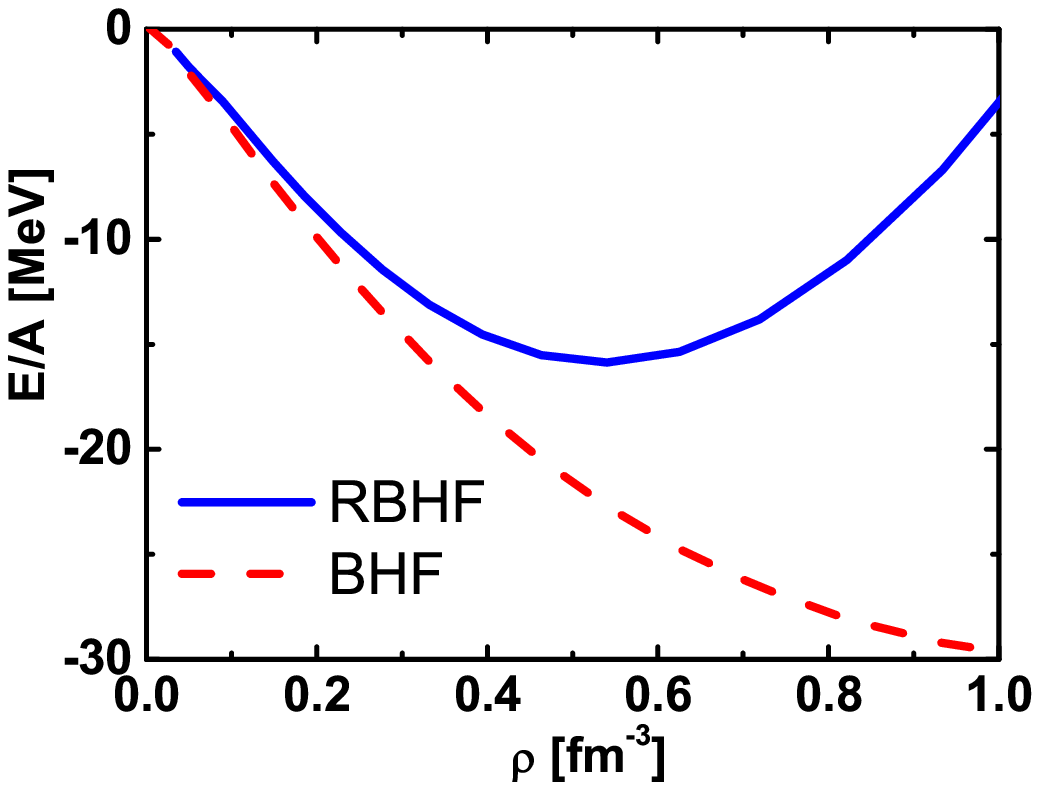}
\caption{The EOSs of symmetric matter with BHF, and RBHF theories with LOBEP1 potential included $P$-waves ($L=1$ channels) contribution. The BHF results are shown in dashed curve. The RBHF ones are given in solid line. }
\label{EOSSDP}
\end{figure}

Now, we can obtain the $P-$wave components of the LOBEP1 potential, which can be regarded as our prediction on the lattice $NN$ potential. Then, this $P-$wave contribution is taken into account in the calculation of nuclear matter. With its contribution, the saturation energy in BHF becomes larger than the one in the RBHF theory now as shown in Fig.\,\ref{EOSSDP} and consistent with the RBHF theory to Bonn potential. It is demonstrated that the $P-$waves are very important for the relativistic effect to provide the repulsive effect. For the LOBEP2 potential, there is a similar result.

\begin{figure}[!htb]
\centering
\includegraphics[bb=0 0 320 240, width=0.50\textwidth]{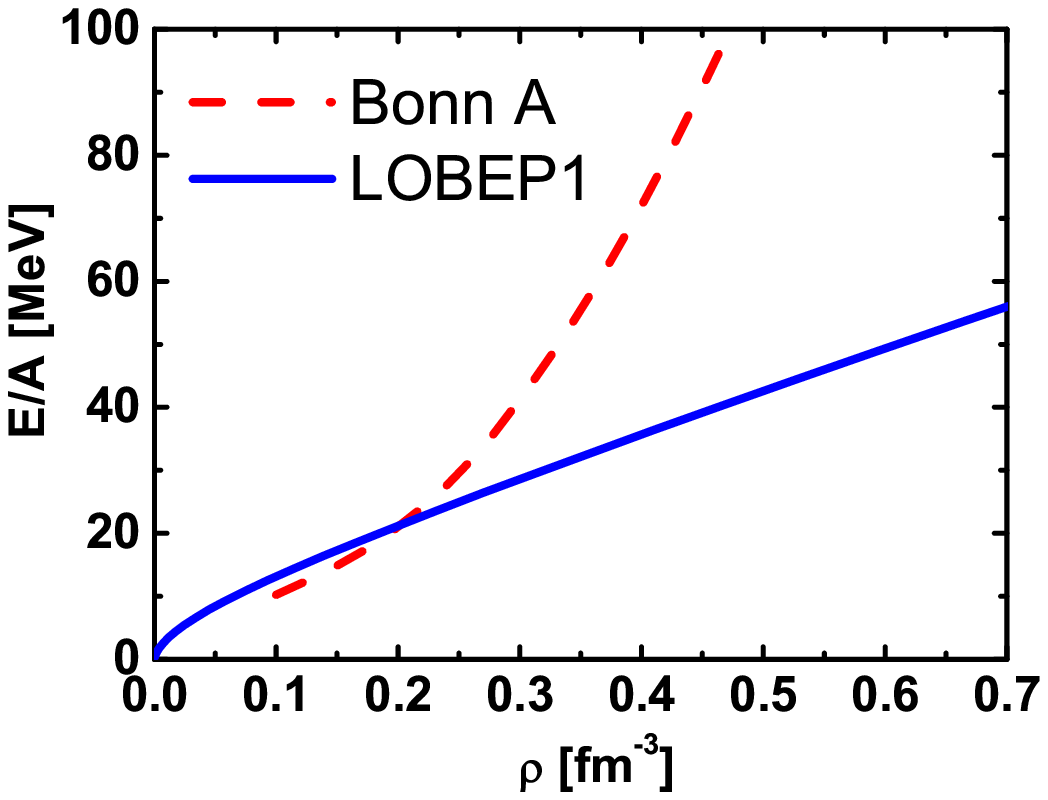}
\caption{The EOSs of pure neutron matter with LOBEP1 and Bonn\,A potential in RBHF theory. The Bonn\,A potential results are shown in dashed curve. The LOBEP1 ones are given in solid line.}
\label{EOSC}
\end{figure}

We also compare the EOSs of pure neutron matter with LOBEP1 and Bonn\,A potentials in the RBHF theory in Fig.\,\ref{EOSC}.  The EOS of Bonn\,A potential has more repulsive effect in the high density region, which includes the contribution not only from $^1S_0$ channel, but also from other $T=1, L>2$ channels, such as $^3P_1$ channel and so on. Therefore, we need more data of lattice $NN$ potential for higher partial waves to describe the properties of neutron star correctly.

\section{Conclusion}
In conclusion, two kinds of one-boson-exchange potential (OBEP) were constructed based on the latest lattice $NN$ potential (La469)  to study the properties of nuclear matter with the relativistic Brueckner-Hartree-Fock (RBHF) theory. We fitted the OBEP with the on-shell matrix elements and phase shifts of La469 potential respectively, and obtained LOBEP1 and LOBEP2, which could completely reproduce the fitting data. The saturation properties of these two OBEPs in the BHF theory were consistent with the existing calculation for La469 potential directly including $^1S_0,~^3S_1,~^3S_1-{^3D_1},$ and $^3D_1$ channels by Inoue {\it{et al.}}. The non-relativistic EOS had more attractive contribution to the saturation energy than the RBHF theory. This result is opposite in comparison with the previous calculation in RBHF theory with Bonn potential, which is obtained by the nucleon-nucleon scattering data. This is because in our calculation, there are only $L=0, ~2$ channels provided by the La469 potential.  Once the $P$-waves with $L=1$ included in LOBEP1 and LOBEP2 potentials, the EOSs of BHF theory became more attractive than the one of RBHF theory, which is consistent with the previous calculation with Bonn potential. It demonstrated that the $P$-waves were very important for the repulsive contributions in relativistic effect. The pure neutron matter is also calculated in BHF and RBHF theory. These EOSs are almost identical from both LOBEP1 potential and LOBEP2 potential. In this case, there are only $^1S_0$ channel contributing to the isospin $T=1$ system.

Although we can obtain the binding state of symmetric nuclear matter with present lattice potential, the saturation properties were still far from the empirical data. The pure neutron matter with lattice potential still need more repulsive contribution at high density to obtain the reasonable maximum mass of neutron star. We hope that the lattice QCD can provide more data on higher partial waves in the $NN$ potential and less quark mass to approach the physical pion mass so that we can realize the {\it{ab initio}} calculation of nuclear many-body system from the QCD level.

\section{Acknowledgments}
J. Hu would like to thank Dr. T. Inoue for providing the subroutine of lattice $NN$ potential and useful discussion and Dr. Y. Zhang for the corrections on language. This work was supported in part by the National Natural Science Foundation of China (Grants No. 11375089, and No. 11405090).

\end{document}